\documentclass{elsart}

\usepackage{amsmath,amsfonts,amssymb}
\usepackage{array,delarray}
\usepackage[dvips]{graphicx}

\usepackage{dsfont}
\usepackage[square]{natbib}

\newcommand{\ff}{\mathds{F}}

\newcommand{\XOR}{\mathrm{XOR}}
\newcommand{\band}{\mathrm{AND}}
\newcommand{\OR}{\mathrm{OR}}
\newcommand{\NOT}{\mathrm{NOT}}
\newcommand{\ncf}{\mathrm{ncf}}
\newcommand{\id}{\mathrm{id}}

\newtheorem{theorem}{Theorem}[section]   
\newtheorem{corollary}[theorem]{Corollary}     
\newtheorem{lemma}[theorem]{Lemma}         

\theoremstyle{definition}
\newtheorem{definition}[theorem]{Definition}   

\theoremstyle{remark}
\newtheorem{remark}[theorem]{Remark}        
\newtheorem{example}[theorem]{Example}        

\numberwithin{equation}{section}     

\def\ds {\displaystyle}

\begin{document}

\begin{frontmatter}

\title{Nested Canalyzing, Unate Cascade, and Polynomial Functions\thanksref{reviewer}}

\author{Abdul Salam Jarrah\corauthref{cor}\thanksref{vbi}}
\ead{ajarrah@vbi.vt.edu}
\corauth[cor]{The corresponding author.}
\author{Blessilda Raposa\thanksref{bpadd}}
\ead{raposab@dlsu.edu.ph}
\author{Reinhard Laubenbacher\thanksref{vbi}}
\ead{reinhard@vbi.vt.edu}

\thanks[reviewer]{
The authors thank the referee for the constructive and insightful
comments which have improved this paper significantly.
}

\address[vbi]{Virginia Bioinformatics Institute (0477), Virginia Tech, Blacksburg, VA 24061, USA}
\address[bpadd]{Mathematics Department, De La Salle University, 2401 Taft Avenue, Manila, Philippines}

\begin{abstract}
This paper focuses on the study of certain classes of Boolean
functions that have appeared in several different contexts. Nested
canalyzing functions have been studied recently in the context of
Boolean network models of gene regulatory networks. In the same
context, polynomial functions over finite fields have been used to
develop network inference methods for gene regulatory networks.
Finally, unate cascade functions have been studied in the design of
logic circuits and binary decision diagrams.  This paper shows that
the class of nested canalyzing functions is equal to that of unate
cascade functions.  Furthermore, it provides a description of nested
canalyzing functions as a certain type of Boolean polynomial
function.  Using the polynomial framework one can show that the
class of nested canalyzing functions, or, equivalently, the class of
unate cascade functions, forms an algebraic variety which makes
their analysis amenable to the use of techniques from algebraic
geometry and computational algebra.  As a corollary of the
functional equivalence derived here, a formula in the literature for
the number of unate cascade functions provides such a formula for
the number of nested canalyzing functions.
\end{abstract}


\end{frontmatter}

\section{Introduction}

Canalyzing functions were introduced by Kauffman \cite{Kauff1} as
appropriate rules in Boolean network models of gene regulatory
networks.  The definition is reminiscent of the concept of
``canalisation" introduced by the geneticist Waddington \cite{wad}
to represent the ability of a genotype to produce the same phenotype
regardless of environmental variability. Canalyzing functions are
known to have other important applications in physics, engineering
and biology. They have been used to study the convergence behavior
of a class of nonlinear digital filters, called stack filters, which
have applications in image and video processing \cite{gabbouj,
wendt, yu}. Canalyzing functions also play an important role in the
study of random Boolean networks \cite{Kauff1, lynch, stauffer,
stern}, have been used extensively as models for dynamical systems
as varied as gene regulatory networks \cite{Kauff1}, evolution,
\cite{stern} and chaos \cite{lynch}.  One important characteristic
of canalyzing functions is that they exhibit a stabilizing effect on
the dynamics of a system. For example, in \cite{gabbouj}, it is
shown that stack filters which are defined by canalyzing functions
converge to a fixed point called a root signal after a finite number
of passes. Moreira and Amaral  \cite{moreira}, showed that the
dynamics of a Boolean network which operates according to canalyzing
rules is robust with regard to small perturbations.

A special type of canalyzing function, so-called \emph{nested
canalyzing functions} (NCFs) were introduced recently in
\cite{Kauff2}, and it was shown in \cite{Kauff} that Boolean
networks made from such functions show stable dynamic behavior and
might be a good class of functions to express regulatory
relationships in biochemical networks. Little is known about this
class of functions, however.  For instance, there is no known
formula for the number of nested canalyzing functions in a given
number of variables.

Another field in which special families of Boolean functions have
been studied extensively is the theory of computing, in particular
the design of efficient logical switching circuits. Since the 1970s,
several families of Boolean functions have been investigated for use
in circuit design.  For instance, the family of {\it fanout-free}
functions has been studied extensively, as well as the family of
cascade functions. A subclass of these are the {\it unate cascade
functions} see, e.g., \cite{maitra, mukho}, which we focus on here.
It turns out that this class of functions has some very useful
properties.  For instance, it was shown recently \cite{butler} that
the class of unate cascade functions is precisely the class of
Boolean functions that have good properties as binary decision
diagrams.  In particular, the unate cascade functions (on $n$
variables) are precisely those functions whose binary decision
diagrams have the smallest average path length $\ds
(2-\frac{1}{2^{n-1}})$ among all Boolean functions of $n$ variables.

The notion of average path length is one cost measure for binary
decision trees, which measures the average number of steps to
evaluate the function on which the tree is based.  One way of
assessing the relative efficacy of classes of Boolean function for
logic circuit or binary decision tree design is to look at the
number of different circuits or trees that can be realized with a
particular class.  That is, one would like to count the number of
functions in a given class.  This has led to a formula for the
number of unate cascade functions \cite{bendbut}. One of the results
in this paper shows that the classes of unate cascade functions and
nested canalyzing functions are identical (as classes of functions
rather than as classes of logical expressions).  As a result of the
equivalence we will establish, this formula then also counts the
number of nested canalyzing functions.

A third framework for studying Boolean functions, in the context of
models for biochemical networks, was introduced in \cite{LS}. There,
a new method to reverse engineer gene regulatory networks from
experimental data was proposed. The proposed modeling framework is
that of time-discrete deterministic dynamical systems with a finite
set of states for each of the variables. The number of states is
chosen so as to support the structure of a finite field. One
consequence is that each of the state transition functions can be
represented by a polynomial function with coefficients in the finite
field, thereby making available powerful computational tools from
polynomial algebra. This class of dynamical systems in particular
includes Boolean networks, when network nodes take on two states. It
is straightforward to translate Boolean functions into polynomial
form, with multiplication corresponding to AND, addition to XOR, and
addition of the constant 1 to negation.  In this paper we provide a
characterization of those polynomial functions over the field with
two elements that correspond to nested canalyzing (and, therefore,
unate cascade) functions.  Using a parameterized polynomial
representation, one can characterize the parameter set in terms of a
well-understood mathematical object, a common method in mathematics.
This is done using the concepts and language from algebraic
geometry. To be precise, we describe the parameter set as an
algebraic variety, that is a set of points in an affine space that
represents the set of solutions of a system of polynomial equations.
This algebraic variety turns out to have special structure that can
be used to study the class of nested canalyzing functions as a rich
mathematical object.

\section{Boolean Nested Canalyzing and unate cascade Functions are equivalent}

\subsection{Boolean Nested Canalyzing Functions}

Boolean nested canalyzing functions were introduced recently in
\cite{Kauff2}, and it was shown in \cite{Kauff} that Boolean
networks made from such functions show stable dynamic behavior. In
this section we show that the set of Boolean nested canalyzing
functions is equivalent to the set of unate cascade functions that
has been studied before in the engineering and computer science
literature. In particular, this equivalence provides a formula for
the number of nested canalyzing functions in a given number of
variables.

We begin by defining the canalyzing property.
\begin{definition}
A Boolean function $f(x_1,\ldots ,x_n)$ is \emph{canalyzing} if
there exists an index $i$ and a Boolean value $a$ for $x_i$ such
that $f(x_1, \ldots ,x_{i-1},a,x_{i+1},\ldots ,x_n) = b$ is
constant. That is, the variable $x_i$, when given the
\emph{canalyzing value} $a$, determines the value of the function
$f$, regardless of the other inputs.  The output value $b$ is called
the \emph{canalyzed value}.
\end{definition}

Throughout this paper, we use the Boolean functions $\band(x,y) = x
\wedge y$, $\OR(x,y) = x\vee y$ and $\NOT(x) = \overline{x}$.
\begin{example}
The function $\band(x,y) = x \wedge y$ is a canalyzing function in
the variable $x$ with canalyzing value 0 and canalyzed value 0. The
function $\XOR(x,y) := (x\vee y)\wedge \overline{(x\wedge y)}$ is
not canalyzing in either variable.
\end{example}

Nested canalyzing functions are a natural specialization of
canalyzing functions. They arise from the question of what happens
when the function does not get the canalyzing value as input but
instead has to rely on its other inputs. Throughout this paper, when
we refer to a function of $n$ variables, we mean that $f$ depends on
all $n$ variables.  That is, for $1 \leq i \leq n$, there exists
$(a_1,\dots,a_n) \in \ff_2^n$ such that
$f(a_1,\dots,a_{i-1},a_i,a_{i+1},\dots,a_n) \neq
f(a_1,\dots,a_{i-1},\overline{a_i},a_{i+1},\dots,a_n)$.

\begin{definition}\label{def-ncf}
Let $f$ be a Boolean function in $n$ variables.
\begin{itemize}
\item Let $\sigma$ be a permutation on $\{1,\dots,n\}$.
The function $f$ is  \emph{nested canalyzing function}(NCF) in the
variable order $x_{\sigma(1)},\dots,x_{\sigma(n)}$ with canalyzing
input values $a_1,\dots, a_n$ and canalyzed  output values
$b_1,\dots,b_n$, respectively,  if it can be represented in the form
\begin{equation} \label{ncf kauff}
f(x_1,x_2,\ldots, x_n) =
\begin{cases}
    b_1 & ~{\rm if}~ x_{\sigma(1)} = a_1, \\
    b_2 & ~{\rm if}~ x_{\sigma(1)} \ne a_1 ~{\rm and}~ x_{\sigma(2)} = a_2, \\
    b_3 & ~{\rm if}~ x_{\sigma(1)} \ne a_1 ~{\rm and}~ x_{\sigma(2)} \ne a_2 ~{\rm and}~ x_{\sigma(3)} = a_3, \\
    \vdots & \hspace{1cm} \vdots \\
    b_n & ~{\rm if}~ x_{\sigma(1)} \ne a_1 ~{\rm and}~  \cdots ~{\rm and}~ x_{\sigma(n-1)} \ne a_{n-1} ~{\rm and}~ x_{\sigma(n)} = a_n, \\
    \overline{b_{n}} & ~{\rm if}~ x_{\sigma(1)} \ne a_1 ~{\rm and}~ \cdots ~{\rm and}~ x_{\sigma(n)} \ne a_n.
\end{cases}
\end{equation}
\item
The function $f$ is nested canalyzing if $f$ is nested canalyzing in
the variable order $x_{\sigma(1)},\dots,x_{\sigma(n)}$ for some
permutation $\sigma$.
\end{itemize}
\end{definition}

\begin{example}
The function $f(x,y,z) = x\wedge \overline{y}\wedge z$ is nested
canalyzing in the variable order $x,y,z$ with canalyzing values
0,1,0 and canalyzed values 0,0,0, respectively. However, the
function $f(x,y,z,w) = x\wedge y\wedge \XOR(z,w)$ is not nested
canalyzing because if $x =1$ and $y =1$, then the value of the
function is not constant for any input values for either $z$ or $w$.
\end{example}
The following lemma follows directly from the definition above.
\begin{lemma}\label{referee}
A Boolean function $f$ on $n$ variables is nested canalyzing  in the
variable order $x_{\sigma(1)},\dots,x_{\sigma(n)}$ with canalyzing
input values $a_1,\dots, a_n$ and canalyzed  output values
$b_1,\dots,b_n$, respectively if and only if $f$ depends on all $n$
variables and, for all $ 1 \leq i \leq n$,
\[
f(c_1,\dots,c_n) = b_i,
\]
where $(c_1,\dots,c_n) \in \ff_2^n$ such that $c_{\sigma(i)} = a_i$
and, for $1 \leq j < i$, $c_{\sigma(j)} = \overline{a_j}$.
\end{lemma}
The next lemma gives the functional form of a nested canalyzing
function.  To simplify notation, we will use the following
notational convention.  Let $a=0,1$.  Then $x+a$ will denote $x$ if
$a=0$ and $\overline{x}$ if $a=1$.
\begin{lemma}\label{ncf-unate}
Let
\begin{equation}\label{unate-eqn}
g(x_1,\dots,x_n) =
(x_{\sigma(1)}+a_1+b_1)\Diamond_1((x_{\sigma(2)}+a_2+b_2)\Diamond_2
(\cdots((x_{\sigma(n-1)}+a_{n-1}+b_{n-1})\Diamond_{n-1}(x_{\sigma(n)}+a_n+b_n))\cdots),
\end{equation}
where
\begin{equation}\label{diamond}
\Diamond_i = \left\{%
            \begin{array}{ll}
             \vee , & \hbox{ if }  b_i =1;\\
             \wedge , & \hbox{ if }  b_i = 0,\\
            \end{array}%
        \right.
\end{equation}
and $a_i,b_i \in \{0,1\}$  for all $1 \leq i \leq n$. Then $g$ is
nested canalyzing in the variable order
$x_{\sigma(1)},\dots,x_{\sigma(n)}$ with canalyzing  input values
$a_1,\dots, a_n$ and canalyzed  output values $b_1,\dots,b_n$,
respectively. Furthermore, any nested canalyzing function can be
represented in the form {\rm (\ref{unate-eqn})}.
\end{lemma}

\begin{pf}
It is clear that $g$ depends on all variables $x_1,\dots, x_n$. Let
$g_n = x_{\sigma(n)}+a_n+b_n$ and, for $1 \leq i < n$, let
\[
g_i = (x_{\sigma(i)}+a_{i}+b_{i}) \Diamond_{i} g_{i+1}.
\]
Then $g = g_1 = (x_{\sigma(1)}+a_1+b_1) \Diamond_1 g_2$.

If $x_{\sigma(1)} = a_1$, then $(x_{\sigma(1)}+a_1+b_1) \Diamond_1
g_2=b_1\Diamond_1 g_2=b_1$, by equation (\ref{diamond}). For $ 1
\leq i \leq n-1$, suppose $x_{\sigma(j)}=\overline{a_j}$ for $j < i$
and $x_{\sigma(i)} = a_{i}$. Now, for all $j < i$, we have
$\overline{b_{j}} \Diamond_{j} g_{j+1}= g_{j+1}$ and
$b_{i}\Diamond_{i} g_{i+1} = b_{i}$. Thus, by equation
(\ref{diamond}), we get
\[
\overline{b_1}\Diamond_1 (\overline{b_2}\Diamond_2(\dots
(b_{i}\Diamond_{i} g_{i+1})\dots)) = b_{i}\Diamond_{i} g_{i+1} =
b_{i}.
\]
Hence $g$ is nested canalyzing, with the $a_i$ as canalyzing values
and the $b_i$ as canalyzed values. It is left to show that any
nested canalyzing function can be represented in the form
(\ref{unate-eqn}). Let $f$ be a nested canalyzing function in the
variable order $x_{\sigma(1)},\dots,x_{\sigma(n)}$ with canalyzing
input values $a_1,\dots, a_n$ and canalyzed output values
$b_1,\dots,b_n$, respectively. By Lemma \ref{referee} and the above,
it is clear that $f(c_1,\dots,c_n)=g(c_1,\dots,c_n)$ for all
$(c_1,\dots,c_n) \in \ff_2^n$.
 Thus $f=g$ as functions and hence $f$ can be represented in the form (\ref{unate-eqn}).
\end{pf}

\subsection{NCFs are Unate Cascade Functions}\label{unate}

We next show that Boolean NCFs are equivalent to unate cascade
functions. Unate cascade functions have been defined and studied
\cite{maitra, mukho} as a special class of fanout-free functions
which are used in the design and synthesis of logic circuits and
switching theory \cite{butler,hayes}.

\begin{definition}
A Boolean function $f$ is a \emph{unate cascade} function if it can
be represented as
\begin{equation} \label{unate casc}
 f(x_1,x_2,\ldots, x_n) = x_{\sigma(1)}^\ast \Diamond_1( x_{\sigma(2)}^\ast
 \Diamond_2(\ldots (x_{\sigma(n-1)}^\ast \Diamond_{n-1} x_{\sigma(n)}^\ast)) \ldots ),
\end{equation}
where $\sigma$ is a permutation on $\{1,\dots,n\}, \, x^\ast$ is
either $x$ or $x + 1$ and $\Diamond_i$ is either the OR ($\vee$) or
AND ($\wedge$) Boolean operator.
\end{definition}

\begin{theorem} \label{ncf = uc}
A Boolean function is nested canalyzing if and only if it is a unate
cascade function.
\end{theorem}

\begin{pf}
Let $f$ be a unate cascade function in the form (\ref{unate casc}).
Let $a_n$ and $b_n$ be such that $x_{\sigma(n)}^\ast =
x_{\sigma(n)}+a_n+b_n$ and, for $1 \leq i < n$, let
\begin{equation}\label{bs}
b_i =   \left\{%
            \begin{array}{ll}
             1 , & \hbox{ if }  \Diamond_i = \vee;\\
             0 , & \hbox{ if }  \Diamond_i = \wedge,\\
            \end{array}%
        \right.
\end{equation}
and let $a_{i} \in \{0,1\}$ such that $x_{\sigma(i)}^\ast =
x_{\sigma(i)}+a_i+b_i$. That is,
\begin{equation}\label{ais}
a_i = x_{\sigma(i)}^\ast + x_{\sigma(i)} + b_i.
\end{equation}
Then
\[
f(x_1,\dots,x_n) =
(x_{\sigma(1)}+a_1+b_1)\Diamond_1((x_{\sigma(2)}+a_2+b_2)\Diamond_2(\cdots
((x_{\sigma(n-1)}+a_{n-1}+b_{n-1})
\Diamond_{n-1}(x_{\sigma(n)}+a_n+b_n))\cdots),
\]
which is nested canalyzing by Lemma \ref{ncf-unate}.

Conversely, let $f$ be a nested canalyzing function of the form
(\ref{ncf kauff}). By Lemma \ref{ncf-unate} and equation
(\ref{unate-eqn}), $f$ can be represented in the form (\ref{unate
casc}) where $x_{\sigma(i)}^\ast= x_{\sigma(i)}+a_i+b_i$ for all $1
\leq i \leq n$. Thus $f$ is unate cascade.
\end{pf}

\begin{remark}\label{two-sets}
The second sentence in the proof above implies the nested canalyzing
function with canalyzing input $(a_1,\dots,a_n)$ and canalyzed
output $(b_1,\dots,b_n)$ is also nested canalyzing in the same
variable order with canalyzing input $(a_1,\dots,a_{n-1},
\overline{a_n})$ and canalyzing output $(b_1,\dots,b_{n-1},
\overline{b_n})$.
\end{remark}
\begin{remark}
The theorem above provides natural equivalence between the class of
nested canalyzing functions and that of unate cascade functions.
Namely, for a given unate cascade function in the form (\ref{unate
casc}), using (\ref{ais}) and (\ref{bs}) we can explicitly define
the canalyzing input values $a_i$ and canalyzed output values $b_i$.
On the other hand, any nested canalyzing function in the form
(\ref{ncf kauff}) can be presented as a unate cascade function in
the form (\ref{unate casc}) where $x_{\sigma(i)}^\ast=
x_{\sigma(i)}+a_i+b_i$ and $\Diamond_{i}$ as in (\ref{diamond}) for
all $1 \leq i \leq n$.
\end{remark}

Using the theorem and remark above,
 it is now possible to translate results about one type
of function into results about the other type.  We point out one
such example.  Several papers have been dedicated to counting the
number of certain fanout-free functions including the unate cascade
functions \cite{bender, bendbut, hayes, pogosyan, sasao}. On the
other hand, Just et. al. \cite{just}, gave a formula for the number
of canalyzing functions.

Bender and Butler \cite{bendbut} and Sasao and Kinoshita
\cite{sasao} independently  found the number of unate cascade
functions, among other fanout-free functions. As an immediate
corollary of Theorem \ref{ncf = uc}, we therefore know the number of
NCFs in $n$ variables, for a given value of $n$. We use the
recursive formula found by Sasao and Kinoshita \cite{sasao}, in the
following corollary.
\begin{corollary}
The number of NCFs in $n$ variables, denoted by $NCF(n)$, is given
by
\[
NCF(n) = 2 \cdot E(n),
\]
where
\[
E(1) = 1, ~~E(2) = 4,
\]
and, for $n \geq 3$,
\[
E(n) = \ds \sum_{r = 2}^{n-1} {n \choose r-1} \cdot 2^{r-1} \cdot
E(n - r + 1) + 2^n.
\]
\end{corollary}

For example, the number of Boolean NCFs (unate cascade functions) on
$n$ variables  for $n \le 8$ is given by Table \ref{unate-tab},
which is part of the tables given by Sasao and  Kinoshita
\cite{sasao} and also Bender and Butler \cite{bendbut}.

\begin{table}[hbtp]
  \begin{center}
  \bigskip
  \caption{\label{unate-tab}
     The number of NCFs on $n \le 8$ variables}      
  \medskip
  \begin{tabular}{|c|c|c|c|c|c|c|c|c|c|c|} \hline
 $n$ & 1 & 2 & 3 & 4 & 5 & 6 & 7 & 8\\ \hline
  $NCF(n)$   & 2& 8 & 64 & 736 & 10,624 & 183,936 & 3,715,072 & 85,755,392\\\hline
  \end{tabular}
  \end{center}
\end{table}

Some interesting facts derive from the equivalence of
 NCFs with unate cascade functions. Sasao and Kinoshita
\cite[Lemma 4.1]{sasao}, found that unate cascade functions are
equivalent to fanout-free threshold functions. Thus NCFs are a
special class of threshold functions. It is also interesting to note
that among switching networks, the unate cascade functions are
precisely those with the smallest average path length, as shown by
Butler et. al. \cite{butler}, which makes them efficient in logic
circuit design.

\section{Nested canalyzing functions as polynomial functions}

Wanting to compute the total number of Boolean functions of a
particular type, e.g. nested canalyzing or unate cascade functions,
is one example of the need to study the totality of such functions.
Few tools other than elementary combinatorics are available for this
purpose, however.  In this section, we propose an alternative
approach to Boolean functions which provides a whole new set of
mathematical tools and results.  We will view Boolean functions
$\{0, 1\}^n\longrightarrow \{0, 1\}$ as polynomial functions $f:
\ff_2^n\longrightarrow \ff_2$, where $\ff_2$ denotes the field with
two elements.  It is well-known that any function $f:
\ff_2^n\longrightarrow \ff_2$ can be represented as a polynomial
function \cite{LN}.  If we require that every variable appear with
exponent 1, then this representation is unique.  For Boolean
functions, this representation is straightforward to construct by
observing that $\band(x,y)=x \wedge y = xy$, $\OR(x,y)= x\vee y
=x+y+xy$, and $\NOT(x)=\overline{x}=x+1$. Conversely, replacing
multiplication by AND and addition by the XOR function, we can
translate any polynomial function into a Boolean function.  In
particular, this shows that any binary function on $n$ variables can
be represented as a Boolean function.  While this seems like a
simple change of language, it has the profound effect of placing the
study of Boolean functions into the fields of algebra and algebraic
geometry, which have a rich body of results and algorithms
available.

The goal of this section is to formalize this equivalence and to
characterize those polynomial functions that represent nested
canalyzing functions. The characterization will be expressed as a
parametrization, with the set of parameters taken from an algebraic
variety.  Algebraic geometry has many tools to study varieties, an
approach that will be pursued elsewhere.

\subsection{Polynomial form of nested canalyzing functions}

We derive a polynomial representation of the class of Boolean nested
canalyzing functions which we then use to identify necessary and
sufficient relations among their coefficients.

Any Boolean function in $n$ variables is a map $f: \{0,1\}^n
\longrightarrow \{0,1\}$. The set of all such maps, denoted by
$B_n$, can be given the algebraic structure of a ring with the
boolean operators $\XOR$ for addition and the conjunction $\band$ for
multiplication.

Consider the polynomial ring $\ff_2[x_1,\dots,x_n]$ over the field
$\ff_2 := \{0,1\}$ with two elements. Let $I$ be the ideal generated
by the polynomials $x_i^2-x_i$ for all $i=1,\dots,n$. (That is, $I$
consists of all linear combinations of these polynomials with
arbitrary polynomials as coefficients.)  For any Boolean function $f
\in B_n$, there is a unique polynomial $g \in \ff_2[x_1,\dots,x_n]$
such that $g(a_1,\dots,a_n) = f(a_1,\dots,a_n)$ for all
$(a_1,\dots,a_n) \in \ff_2^n$ and such that the degree of each
variable appearing in $g$ is equal to $1$. Namely
\begin{equation}\label{poly-rep}
g(x_1,\dots,x_n) = \sum_{(a_1,\dots,a_n) \in \ff_2^n}
f(a_1,\dots,a_n) \prod_{i=1}^n (1-(x_i-a_i)).
\end{equation}
And it is straightforward to show that this equivalence extends to a
ring isomorphism
$$
R := \ff_2[x_1,\dots,x_n]/I \cong B_n.
$$  From now on we will not distinguish between these two rings.

Next we present and study the set of all Boolean nested canalyzing
functions as a subset of the ring $R$ of all Boolean polynomial
functions.

The following theorem gives the polynomial form for canalyzing and
nested canalyzing functions.

\begin{theorem}\label{ncf poly}
Let $f$ be a  function in $R$. Then
\begin{enumerate}
\item The function $f$ is  \emph{canalyzing} in the variable $x_i$, for some $i, ~~ 1 \le i \le n$, with
    canalyzing input value $a_i$ and canalyzed output value $b_i$, if and only if
    \begin{equation}\label{can}
    f(x_1,x_2, \ldots, x_i, \ldots, x_n) = (x_i -a_i) g(x_1, x_2, \ldots, x_i, \ldots, x_n) + b_i.
    \end{equation}
\item The function $f$ is  \emph{nested canalyzing}  in the order $x_1, x_2, \ldots, x_n$, with canalyzing
    values $a_i$ and corresponding canalyzed values $b_i$, $1 \le i \le n$, if and only if it
    has the polynomial form
    \begin{equation} \label{ncf q factor}
    \begin{array}{lll}
    f(x_1,x_2,\ldots, x_n) &=& (x_1-a_1) [(x_2 - a_2)[\ldots [ (x_{n-1} - a_{n-1}) [ (x_n-a_n)  \\
          && +( b_n-b_{n-1}) ] + (b_{n-1}-b_{n-2}) ]   \ldots ]+(b_2-b_1)]  + b_1 \\
    \end{array}
    \end{equation}
    or, equivalently,
    \begin{equation} \label{ncf q}
    f(x_1,x_2,\ldots, x_n) =  \ds  \prod_{i=1}^n (x_i- a_i) + \sum_{j=1}^{n-1}\left [ (b_{n-j+1} - b_{n-j}) \prod_{i=1}^{n-j}(x_i-a_i) \right] + b_1.
    \end{equation}
\end{enumerate}
\end{theorem}

\begin{pf} \hfill
\begin{enumerate}
\item It is easy to see that if $x_i = a_i$, then the output is $b_i$,
    no matter what the values of the other variables are.
    Conversely, if $f$ is canalyzing with input $a_i$ and output
    $b_i$, then $f-b_i$, as a polynomial in $x_i$ has $a_i$ as a root,
    hence is divisible by $x_i-a_i$.  This proves
    the first claim.
\item Let $f$ be a nested canalyzing function as in Definition
\ref{def-ncf}, and let
\begin{equation*}
\begin{array}{lll}
g(x_1,\dots,x_n) &=& (x_1-a_1) [(x_2 - a_2)[\ldots [ (x_{n-1} - a_{n-1}) [ (x_n-a_n)  \\
      && +( b_n-b_{n-1}) ] + (b_{n-1}-b_{n-2}) ]   \ldots ]+(b_2-b_1)]  + b_1. \\
    \end{array}
\end{equation*}
We will show that $g$ is the unique polynomial presentation of $f$,
as in equation \ref{poly-rep}. Since the degree of each variable in
$g$ is equal to one, we only need to show that $g(c_1,\dots,c_n) =
f(c_1,\dots,c_n)$ for all $(c_1,\dots,c_n) \in \ff_2^n$.

Clearly, if $c_1 = a_1$, then  $g(c_1,\dots,c_n) = b_1$ . If $c_1
\ne a_1$ and $c_2 = a_2$, then $(c_1-a_1)=1$ and $g(c_1,\dots,c_n)
=b_2$. If $c_1 \ne a_1$, $c_2 \ne a_2$ and $c_3 = a_3$, then
$g(c_1,\dots,c_n) = b_3$. We continue until we have $c_i \ne a_i$
for all $1 \leq i < n$ and $c_n= a_n$, in which case we get
$g(c_1,\dots,c_n) = b_n$. If $c_i \ne a_i$ for all $i$, then
$(c_i-a_i)=1$ for all $i$ and hence  $g(c_1,\dots,c_n) = 1+b_n$.
Thus $g$ is the unique polynomial representation of $f$.
\end{enumerate}
\end{pf}

\subsection{A Parametrization of NCFs} \label{coeff}

Our next goal is to derive a criterion as to when a given Boolean
polynomial in $n$ variables is a nested canalyzing function.  The
criterion will be given in terms of a parametrization of such
polynomials corresponding to points in the affine space
$\ff_2^{2^n}$ that satisfy a certain collection of polynomial
equations.  Such a set is by definition an algebraic variety, in the
language of algebraic geometry.  This parametrization describes the
entire space of nested canalyzing functions as a geometric object,
whose properties can then be studied with the tools of algebraic
geometry.

Recall that the ring of Boolean functions is isomorphic to the
quotient ring $R = \ff_2[x_1,\dots,x_n]/I$, where $I = \langle
x_i^2-x_i : 1 \leq i \leq n \rangle$. Therefore, the terms of a
Boolean polynomial consist of square-free monomials. Thus, we can
uniquely index monomials by the subsets of $[n]:=\{1,\ldots ,n\}$
corresponding to the variables appearing in the monomial, so that we
can write the elements of $R$ as
\begin{equation}
R = \{\ds \sum_{S \subseteq [n]} c_S \prod_{i \in S} x_i \, : \, c_S
\in \ff_2\}.
\end{equation}
As a vector space over $\ff_2$, $R$ is isomorphic to $\ff_2^{2^n}$
via the correspondence
\begin{equation}\label{corresp}
R \ni \ds \sum_{S \subseteq [n]} c_S \prod_{i \in S} x_i
\longleftrightarrow (c_{\emptyset},\dots, c_{[n]}) \in \ff_2^{2^n},
\end{equation}
for a given fixed total ordering of all square-free monomials.
 That is, a polynomial function corresponds to the vector of
coefficients of the monomial summands.  In this section we identify
the set of nested canalyzing functions in $R$ with a subset
$V^{\ncf}$ of $\ff_2^{2^n}$ by imposing relations on the coordinates
of its elements.

Let $S$ be any subset of $[n]$. We introduce a new term called the
\emph{completion} of $S$.

\begin{definition}
Let $S$ be a a non-empty set whose highest element is $r_S$. The
{\it completion} of $S$, which we denote by $[r_S]$,  is the set
$[r_S] := \{1,2,\ldots,r_S\}$. For $S = \emptyset$, let
$[r_\emptyset] :=\emptyset$.
\end{definition}

The main result of this section is the following theorem.

\begin{theorem}\label{coeff rel}
Let $f$ be a Boolean polynomial in  $n$ variables, given by
\begin{equation}\label{poly}
f(x_1,x_2,\ldots, x_n) = \sum_{S \subseteq [n]} c_S \prod_{i \in S}
x_i.
\end{equation}
The polynomial $f$ is a nested canalyzing function in the order
$x_1, x_2, \ldots, x_n$ if and only if $c_{[n]} = 1$, and for any
subset $S \subseteq [n]$,
\begin{equation} \label{coeff formula}
c_S = c_{[r_S]} \prod_{i \in [r_S] \backslash S}  c_{[n]\backslash
\{i\}}.
\end{equation}
\end{theorem}

\begin{pf}
First assume that the polynomial $f$ is a Boolean nested canalyzing
function in the order $x_1, x_2, \ldots, x_n$, with canalyzing input
values $a_i$ and corresponding canalyzed output values $b_i, 1\le i
\le n$.  Then, by part 2 of Theorem \ref{ncf poly}, $f$ has the form
(\ref{ncf q}) which can be expanded as
\begin{equation}\label{expanded}
f(x_1,\dots,x_n) = \sum_{S\subseteq [n]} \prod_{i\in S} x_i \prod_{l
\in [n]\backslash S} a_l + \sum_{j=1}^{n-1}(b_{n-j+1}-b_{n-j})
(\sum_{S \subseteq [n-j]} \prod_{i \in S} x_i \prod_{l \in [n-j]
\backslash S} a_l)+b_1.
\end{equation}

We now equate corresponding coefficients in equations (\ref{poly})
and (\ref{expanded}).  First let $S=[n]$.  Then, clearly,
$c_{[r_S]}=1$. Next, consider subscripts of the form
$S=[n]\backslash \{i\}, i\neq n$, that is, coefficients of monomials
of total degree $n-1$ which contain $x_n$. It is clear from equation
(\ref{expanded}) that $x_n$ only appears in the first summand and
hence, for $1 \leq i \leq n-1$,
\begin{equation} \label{ai1}
c_{[n]\backslash\{i\}}=a_i=c_{[r_S]}c_{[n]\backslash\{i\}}.
\end{equation}
It is easy to check that equation (\ref{coeff formula}) holds for
any set $S \subseteq [n]$ such that $S = [r_S]$. By equating the
coefficient of $x_1\cdots x_{r_s}$ in equations (\ref{expanded}) and
(\ref{poly}), we get
\begin{eqnarray*}
c_{S} = c_{[r_S]}   &=& \prod_{i \in [n]\backslash S} a_i + (b_n-b_{n-1}) \prod_{i \in [n-1]\backslash S} a_i+\cdots + (b_{r_S+1} - b_{r_S}) \prod_{i \in [r_S]\backslash S} a_i\\
                    &=& \prod_{i \in [n]\backslash S} a_i + (b_n-b_{n-1}) \prod_{i \in [n-1]\backslash S} a_i+\cdots + (b_{r_S+1} - b_{r_S}),
\end{eqnarray*}
since $\ds \prod_{i \in [r_S]\backslash S} a_i = \prod_{i \in
\emptyset} a_i := 1$, by definition. Now let $S$ be any nonempty
index set. Then
\begin{eqnarray*}
c_S &=& \prod_{i \in [n]\backslash S} a_i + (b_n-b_{n-1}) \prod_{i
\in [n-1]\backslash S} a_i+\cdots + (b_{r_S+1} - b_{r_S})
\prod_{i \in [r_S]\backslash S} a_i\\
      &=& \prod_{i \in [r_S]\backslash S} a_i
      [\prod_{i \in [n]\backslash [r_S]} a_i+(b_n - b_{n-1})
      \prod_{i \in [n-1]\backslash [r_S]} a_i+\cdots+ (b_{r_S+1} - b_{r_S})] \\
      &=& (\prod_{i \in [r_S]\backslash S} a_i ) c_{[r_S]}\\
      &=& c_{[r_S]} \prod_{i \in [r_S]\backslash S} c_{[n]\backslash \{i\}}.
\end{eqnarray*}

This completes the proof that a nested canalyzing polynomial has to
satisfy equation (\ref{coeff formula}).

\vspace{.2cm} \noindent Conversely, suppose that $c_{[n]} = 1$ and
equation (\ref{coeff formula}) holds for the coefficients of the
polynomial $f$ in equation (\ref{poly}).  We need to show that $f$
is nested canalyzing. Using Lemma \ref{referee}, it is enough to
show that $f$ depends on all $n$ variables and
$f(\overline{a_1},\dots,\overline{a_{j-1}},a_j,x_{j+1},\dots,x_n)
=b_j$ for some $a_j,b_j \in \ff_2^n$ and $1\leq j \leq n$. Since
$c_{[n]} = 1$, the monomial $x_1\cdots x_n$ is a summand in $f$ and
hence $f$ depends on all $n$ variables. Now let $1 \leq j \leq n$.
For any $S\subset [n]$ such that  $j\notin S$ and $r_{S} > j$, we
have
\[
c_S=c_{[r_S]}\prod_{i \in [r_S]\backslash S}c_{[n]\backslash \{i\}}
\mbox{ \, \, and \, \, }   c_{S\cup \{j\}}=c_{[r_S]}\prod_{i \in
[r_S] \backslash \{S\cup \{j\}\}}c_{[n]\backslash \{i\}}.
\]
By pairing $c_S$ with $c_{S\cup \{j\}}$ and $c_T$ with $c_{T\cup
\{j\}}$ where $T\subseteq [j-1]$, we rewrite the form (\ref{poly})
into
\begin{equation}\label{new}
f(x_1,\dots,x_n)= \sum_{T\subseteq [j-1]}(x_jc_{T\cup\{j\}}+c_{T})
\prod_{i\in T} x_i + (c_{[n]\backslash \{j\}} +
x_j)\sum_{\substack{S\subset [n] \\ r_S > j \\ j \notin S}}c_{S\cup
\{j\}} \prod_{i \in S} x_i.
\end{equation}
For $1 \leq j \leq n$, let $a_j = c_{[n]\backslash\{j\}}$. Then
\begin{equation}\label{bj}
f(\overline{a_1},\dots,\overline{a_{j-1}},a_j,x_{j+1},\dots,x_n) =
\sum_{T\subseteq [j-1]}(c_{[n]\backslash\{j\}}c_{T\cup\{j\}}+c_{T})
\prod_{i\in T} (1+c_{[n]\backslash\{i\}})
\end{equation}
is a constant which we call $b_j$. Hence, by Lemma \ref{referee},
the function $f$ is nested canalyzing.
\end{pf}

\begin{remark}
Observe that the relations in equation (\ref{coeff formula}) leave
the coefficients $c_\emptyset$ and $c_{[i]}$, for all $1 \leq i <
n$, undetermined, as well as the coefficients $c_S$, where $S$ is
any of the $(n-1)$-element subsets of $[n]$ which include $n$.
Furthermore, a Boolean NCF requires that $c_{[n]} = 1$. Since a
general Boolean polynomial in $n$ variables has $2^n$ coefficients,
equation (\ref{coeff formula}) yields $2^n - 2n$ equations which
have to be satisfied by the coefficients of a Boolean NCF.
\end{remark}

\begin{corollary}\label{variety-id}
The set of points in $\ff_2^{2^n}$ corresponding to coefficient
vectors of nested canalyzing functions in the variable order
$x_1,\dots,x_n$, denoted by $V_{\id}^{\ncf}$, is given by
\begin{equation}
V_{\id}^{\ncf} = \{(c_\emptyset,\dots,c_{[n]}) \in \ff_2^{2^n} :
c_{[n]}=1, \, c_S = c_{[r_S]} \prod_{i \in [r_S] \backslash S}
c_{[n] \backslash \{i\}} \mbox {, for }\, S \subseteq [n]\}.
\end{equation}
\end{corollary}

The following corollary provides surprisingly simple expressions of
the canalyzing input and canalyzed output values in terms of the
coefficients of the polynomial.

\begin{corollary} \label{ai bi}
Let $f$ be a Boolean polynomial given by equation ({\rm
\ref{poly}}). If the polynomial $f$ is a nested canalyzing function
in the order $x_1, x_2, \ldots, x_n$, with input values $a_j$ and
corresponding output values $b_j, 1\le j \le n$,  then
\begin{eqnarray}
 \label{ai} a_j &=& c_{[n]\backslash \{j\}}, \hskip 3.7 true cm {\rm for }~~ 1 \le j \le n-1  \\
 \label{b1} b_1 &=& c_\emptyset + c_1c_{[n]\backslash \{1\}}, \\
 \label{bi} b_{j+1} - b_{j} &=& c_{[j+1]} c_{[n]\backslash \{j+1\}} + c_{[j]}, \hspace{1cm}  \mbox{\rm for } 1 \le j < n-1 \,\, \mbox{\rm and} \\
 \label{bn-an}  b_n - a_n &=& b_{n-1} + c_{[n-1]}~.
\end{eqnarray}
\end{corollary}

\begin{pf}
Equation (\ref{ai}) follows from equation (\ref{ai1}), equation
(\ref{b1}) follows directly from equation (\ref{bj}) when $j=1$. In
equation (\ref{ncf q factor}), we observe that the variable
$x_{n-1}$ appears only in the first and second group of products. In
particular, $c_{[n-1]} = -a_n + b_n -b_{n-1}$, and hence equation
(\ref{bn-an}) follows.

It is left to show  (\ref{bi}). From equation (\ref{bj}),
\begin{eqnarray*}
b_j   &=&   \sum_{T\subseteq [j-1]}(c_{[n]\backslash\{j\}}c_{T\cup\{j\}}+c_{T}) \prod_{i\in T} (1+ c_{[n]\backslash\{i\}}) \\
      &=&   \sum_{T\subseteq [j-1]}(c_{[n]\backslash\{j\}}c_{[j]}\prod_{i \in [j-1]\backslash T} c_{[n]\backslash\{i\}} +c_{T}) \prod_{i\in T} (1+ c_{[n]\backslash\{i\}}) \\
      &=& c_{[n]\backslash\{j\}}c_{[j]}\sum_{T\subseteq [j-1]}c_{[n]\backslash\{j\}}c_{[j]}\prod_{i \in [j-1]\backslash T} c_{[n]\backslash\{i\}} \prod_{i\in T} (1+ c_{[n]\backslash\{i\}})+   \sum_{T\subseteq [j-1]}c_{T} \prod_{i\in T} (1+ c_{[n]\backslash\{i\}}) \\
      &=& c_{[n]\backslash\{j\}}c_{[j]}+\sum_{T\subseteq [j-1]}c_{T}
\prod_{i\in T} (1+ c_{[n]\backslash\{i\}}),
\end{eqnarray*}
since
\begin{displaymath}
\sum_{T\subseteq [j-1]}\prod_{i \in [j-1]\backslash T}
c_{[n]\backslash\{i\}} \prod_{i\in T} (1+ c_{[n]\backslash\{i\}}) =
\prod_{i\in [j-1]}(c_{[n]\backslash\{i\}} +
(1+c_{[n]\backslash\{i\}})) = \prod_{i\in [j-1]} 1 = 1.
\end{displaymath}
Now
\begin{eqnarray*}
b_{j+1} - b_{j}   &=& c_{[n]\backslash\{j+1\}}c_{[j+1]}+\sum_{T\subseteq [j]}c_{T} \prod_{i\in T} (1+ c_{[n]\backslash\{i\}}) - c_{[n]\backslash\{j\}}c_{[j]}-\sum_{T\subseteq [j-1]}c_{T} \prod_{i\in T} (1+ c_{[n]\backslash\{i\}})  \\
                  &=& c_{[n]\backslash\{j+1\}}c_{[j+1]}- c_{[n]\backslash\{j\}}c_{[j]}+\sum_{\substack{T\subseteq [j] \\ j \in T}}c_{T} \prod_{i\in T} (1+ c_{[n]\backslash\{i\}}) \\
                  &=& c_{[n]\backslash\{j+1\}}c_{[j+1]}- c_{[n]\backslash\{j\}}c_{[j]} +\sum_{T\subseteq [j-1]}(1+c_{[n]\backslash\{j\}}) c_{[j]} \prod_{i \in [j-1] \backslash T} c_{[n]\backslash \{i\}} \prod_{i\in T} (1+ c_{[n]\backslash\{i\}})\\
                  &=& c_{[n]\backslash\{j+1\}}c_{[j+1]}- c_{[n]\backslash\{j\}}c_{[j]}+(1+c_{[n]\backslash\{j\}})c_{[j]}\sum_{T\subseteq [j-1]}  \prod_{i \in [j-1] \backslash T} c_{[n]\backslash \{i\}} \prod_{i\in T} (1+ c_{[n]\backslash\{i\}})\\
                  &=& c_{[n]\backslash\{j+1\}}c_{[j+1]}- c_{[n]\backslash\{j\}}c_{[j]}+(1+c_{[n]\backslash\{j\}})c_{[j]} \\
                  &=& c_{[n]\backslash\{j+1\}}c_{[j+1]} + c_{[j]}
\end{eqnarray*}

\end{pf}

\begin{remark}
Equations (\ref{ai})--(\ref{bi}) imply that the input values $a_i$
and output values $b_i$, $1 \le i \le n-1$, are determined uniquely
by the coefficients of the polynomial $f$. Also, equation (\ref{bn-an})
implies that there are two sets of values for $a_n$ and $b_n$ which
will yield the same nested canalyzing function $f$. Using these
facts, we discover Remark \ref{two-sets}.
\end{remark}

\begin{example}
In Table \ref{tabl2}, we give some examples of relationships between
coefficients of  NCFs in $n$ variables, nested in the order
$x_1,x_2, \ldots, x_n$ for some small values of $n$.
\begin{table}[hbtp]
  \begin{center}
  \bigskip
  \caption{\label{tabl2}
  Some examples of relationships between coefficients of  NCFs}      
  \medskip
  \begin{tabular}{|l|l|l|} \hline
  \phantom{xxxx} $n=3$ & \phantom{xxxx}$n=4$ & \phantom{xxxx}$n=5$ \\ \hline
  $c_3 = c_{13}c_{23}c_{123}$ & $c_4 = c_{234}c_{134}c_{124}c_{1234}$ & $c_5 =
  c_{2345}c_{1345}c_{1245}c_{1235}c_{12345}$ \\
  $c_2 = c_{23}c_{12}$ & $c_{13} = c_{134}c_{123}$ & $c_{124} = c_{1245}c_{1234}$ \\
                        & $c_{24} = c_{234}c_{124}c_{1234}$ & $c_{23} = c_{2345}c_{123}$ \\
                        & $c_2 = c_{234}c_{12}$ &   $c_2 = c_{2345}c_{12}$ \\ \hline
  \end{tabular}
  \end{center}
\end{table}
\end{example}

We now extend Theorem \ref{coeff rel} to the general case when the
variables are nested in any given order. For this, we will need to
extend the definition of {\it completion} of a set $S$ with respect
to any permutation of its elements.

\begin{definition}
Let $\sigma$ be a permutation on the elements of the set $[n]$. We
define a new order relation $<_{\sigma}$ on the elements of $[n]$ as
follows: $i <_\sigma j$ if and only if $\sigma^{-1}(i) <
\sigma^{-1}(j)$. Let $S$ be a nonempty subset of $[n]$, say $S =
\{i_1, \dots, i_t\}$. Let $r_S^{\sigma} :=
\max\{\sigma^{-1}(i_1),\dots,\sigma^{-1}(i_t)\}$. The
\emph{completion of S with respect to the permutation $\sigma$},
denoted by $[r_S^{\sigma}]_\sigma$, is the set
$[r_S^{\sigma}]_\sigma := \{ \sigma(1),\dots,\sigma(r_S^\sigma)\}$.
\end{definition}

The following corollary is a generalization of Theorem \ref{coeff
rel}. It gives necessary and sufficient relations among the
coefficients of a  NCF whose variables are nested in the order
specified by a permutation $\sigma$ on $[n]$.

\begin{corollary} \label{coeff rel sigma}
Let $f \in R$ and let $\sigma$ be a permutation of the set $[n]$.
The polynomial $f$ is a nested canalyzing function in the order
$x_{\sigma (1)}, \ldots, x_{\sigma (n)}$, with input values
$a_{\sigma (i)}$ and corresponding output values $b_{\sigma (i)}, 1
\le i \le n$, if and only if $c_{[n]} = 1 $ and, for any subset $S
\subseteq [n]$,
\begin{equation} \label{coeff formula sigma}
c_S = c_{[r_S^{\sigma}]_\sigma} \prod_{w \in [r_S^{\sigma}]_\sigma
\backslash S}  c_{[n] \backslash \{w\}}.
\end{equation}
\end{corollary}

\begin{pf}
We follow the same argument as in the proof of Theorem \ref{coeff
rel}, where we impose the order relation $<_{\sigma}$ on  the
elements of $[n]$ and  we replace all occurrences of the subscript
$i$  by $\sigma (i)$ and $[r_S]$  by $[r_S^{\sigma}]_\sigma$.
\end{pf}

\begin{corollary}\label{variety-sigma}
Let $\sigma$ be a permutation on $[n]$. The set of points in
$\ff_2^{2^n}$ corresponding to nested canalyzing functions in the
variable order $x_{\sigma(1)}, \dots, x_{\sigma(n)}$, denoted by
$V_{\sigma}^{\ncf}$, is defined by
\begin{equation}
V_{\sigma}^{\ncf} = \{(c_\emptyset,\dots,c_{[n]}) \in \ff_2^{2^n} :
c_{[n]}=1, \, c_S = c_{[r_S^{\sigma}]_\sigma} \prod_{w \in
[r_S^{\sigma}]_\sigma \backslash S} c_{[n] \backslash \{w\}} \mbox
{, \, for } S \subseteq [n]\}.
\end{equation}
\end{corollary}

The following corollary is an extension of Corollary \ref{ai bi} and
gives the input and output values of a Boolean NCF whose variables
are nested in the order specified by some permutation $\sigma$.

\begin{corollary}
Let $f \in R$ and let $\sigma$ be a permutation of the elements of
the set $[n]$. If  $f$ is a nested canalyzing function in the order
$x_{\sigma (1)}, \ldots, x_{\sigma (n)}$, with input values $a_j$
and corresponding output values $b_j, 1\le j \le n$, then
\begin{eqnarray}
\label{ais } a_j &=& c_{[n] \backslash \{\sigma (j) \}}, \hskip 5.2 true cm {\rm for}~~ 1 \le j \le n-1  \\
 \label{b1s} b_1 &=& c_\emptyset + c_{\sigma(1)}c_{[n]\backslash \{\sigma (1) \}}, \\
 \label{bis} b_{j+1}- b_{j}  &=& c_{[j+1]_\sigma} c_{[n] \backslash \{\sigma (j+1)\}} + c_{[j]_\sigma},
 \hskip .75 true cm {\rm for}~~1 \le j < n-1~~~{\rm and} \\
 \label{bn-ans}  b_{n} - a_{n} &=& b_{n-1} + c_{[n-1]_\sigma}.
\end{eqnarray}

\end{corollary}

\begin{pf}
This follows from Corollary \ref{ai bi}, where we  replace all
occurrences of subscript $j$  by $\sigma (j)$ and $[r]$ by
$[r]_{\sigma}$.
\end{pf}

Recall that the set $V^{\ncf}$ of nested canalyzing functions is the
union of the sets $V_\sigma^{\ncf}$ of canalyzing functions with
respect to a specified variable order. By Corollaries
\ref{variety-id}, \ref{variety-sigma}, and the correspondence
(\ref{corresp}), we have
\begin{eqnarray*}
V^{\ncf} &=& \bigcup_\sigma V_\sigma^{\ncf}.
\end{eqnarray*}
Corollary \ref{coeff rel sigma} is the starting point for a
geometric analysis of the set of all nested canalyzing functions. It
provides a set of equations that have to be satisfied by the
coefficient vectors of the polynomial representations of the
functions.  These coefficient vectors therefore form an algebraic
variety in the space $\ff_2^{2^n}$, which turns out to have very
nice properties.  In particular, it is a so-called toric variety.

\section{Discussion}
Our main contribution in this paper is to connect three different
fields of inquiry into Boolean functions, which were heretofore
apparently unconnected.  The equivalence of nested canalyzing
functions and unate cascade functions relates the electrical
engineering point of view of logic circuits with the dynamic
biological network view, providing a dictionary for results.  The
equivalence of both to a class of polynomial functions brings rich
additional mathematical structure to the study of both.  In
particular, the language and concepts of algebraic geometry and the
rich tool set of computational algebra and algebraic geometry
provides a foundation that imposes a mathematical structure on the
entire class of these functions, which suggests an entirely new way
of studying them.  As an algebraic variety, the class of nested
canalyzing functions has a very special structure, namely that of a
toric variety \cite{fulton}. Toric varieties lie at the interface of
geometry, algebra, and combinatorics and have a rich structure
\cite{sturmfels}. In another paper, we will explore the properties
of the toric varieties in the previous section in more detail.

In particular, our motivation for this study originally was the
desire to give a characterization of nested canalyzing functions as
polynomials, which could be used as part of the model selection
algorithm in \cite{LS}.  That is, we are interested in giving an
efficient criterion which allows our symbolic computation algorithm
to preferentially pick nested canalyzing functions rather than
general polynomials.  The characterization of this class as a toric
variety is the first important step in this direction.

It deserves mention that the connection to unate cascade functions
was discovered in a roundabout way.  We first established the
parametrization of nested canalyzing functions by special
polynomials.  The structure of these polynomials makes it easy to
count how many there are for a given number of variables.  After
carrying out this counting procedure for the first few numbers
resulted in a sequence of integers which we submitted to N. Sloane's
integer sequence database ({\tt
http://www.research.att.com/\~{}njas/sequences/}). One of the matching
sequences was that for the number of unate cascade functions.

\section{Acknowledgements}
Jarrah and Laubenbacher were supported partially by NSF Grant DMS-0511441.
Laubenbacher was also supported partially by NIH Grant RO1
GM068947-01, a joint computational biology initiative between NIH
and NSF. Raposa was supported by a Fulbright research
grant while visiting the Virginia Bioinformatics Institute (VBI)
where this research was conducted. She thanks VBI for the
hospitality during her stay.


\end{document}